\documentstyle[12pt,equation]{article} 
 1                                        
\def\vec#1{\mbox{\protect\boldmath $ #1 $}}                                
                               
\begin{document}                                                              
\begin{center}                                                                
{ \large\bf THE QUANTUM MECHANICAL CURRENT OF THE PAULI EQUATION \\}
\vskip 2cm  
{ Marek Nowakowski \\}                              
Grup de F\'{i}sica 
Te\`orica, Universitat Aut\`onoma de Barcelona, 08193 Bellaterra,
Spain 
\end{center}
\vskip .5cm                             
\begin{abstract}
We argue that the process of constructing the quantum mechanical 
current of the
Pauli equation 
by copying the line of arguments used in the spin-$0$ case,
i.e. the Schr\"{o}dinger equation, is ambiguous. 
We show that a non-relativistic reduction of the relativistic Dirac
four-vector 
current is, however, capable of fully resolving the problem. This analysis
reveals that the non-relativistic current of the Pauli equation 
should include an extra term of the form 
$\vec \nabla \vec \times (\psi^{\dagger}
\vec \sigma \psi)$.  
We present an initial exploration of the potential consequences of this
new 'spin-term' by solving the Pauli equation for crossed magnetic and 
electric fields and calculating the corresponding current.
\end{abstract}     
\newpage
Most of the applications of Quantum Mechanics (QM) can be found in the realm
of the three phases of matter: gases, liquids and solids where a 
non-relativistic quantum mechanical description is fully adequate. This 
shows the importance of non-relativistic QM, even though, in principle, a 
strictly correct treatment should implement relativity as well. The basic
constituents of matter are atoms whose building blocks in turn are nuclei and 
electrons. These last control the 
properties of matter such as chemical bonds and conductivity.
Since electrons are spin-$1/2$ fermions, the non-relativistic wave
equation describing them, namely the Pauli equation, has a somewhat
distinguished position in our understanding of the matter surrounding us.
But whereas in most books on QM a considerable effort is spent on the 
interpretation of the Schr\"{o}dinger equation (i.e. the spin-$0$ wave
equation) in terms of the probability density $\rho$ and the current
$\vec j$, a corresponding discussion of the spin-$1/2$ case 
(the Pauli equation) is 
rarely to be found \cite{one}.
This might have to do with the tacit assumption that the construction of the
spin-$1/2$ current goes along the same line of arguments as in the
spin-$0$ case. Hence one might conclude that, 
apart from 
a trivial replacement of `complex-conjugate' by `hermitian-conjugate', 
there is
conceptually nothing new in the current of the Pauli equation.
This, as will be shown below, is misleading. 
Indeed, in constructing the current
for spin-$1/2$ case 
one can copy all the steps known from the Schr\"{o}dinger case to obtain an 
expression for $\vec j$. However, due to the presence of the spin
this construction is ambiguous, i.e. there exist terms which can be added
to $\vec{j}$ and which do not emerge from the above mentioned construction.
They can be added since they do not spoil the continuity equation, they are 
of first order in the derivative and second order of the wave function
which means that they are of `velocity-type' as are the rest 
of the conventional
terms known from the spin-$0$ case. 
This ambiguity cannot be resolved by means of non-relativistic QM 
alone. Or, to put it differently, this ambiguity only appears
from the point of view of non-relativistic QM. However, the current must
be fixed uniquely as it is an observable.

It is clear that something new is required to solve the 
problem in a satisfactory way. 
As it is often the case in physics a new symmetry imposed on
the system restricts the number of possible terms and can
therefore resolve an otherwise persistent ambiguity. 
The symmetry we have in mind here, is the Lorentz symmetry. 
A relativistic wave equation for spin-1/2 particles is the
Dirac equation which has  
a positive
definite probability density and a continuity equation \cite{two}.
The non-relativistic
reduction of the Dirac equation and of the corresponding current
will then answer the problem unambiguously. 
From a pedagogical point of 
view it seems even desirable to postpone the discussion of the Pauli 
current until relativistic QM is introduced. Unfortunately, the main stream of
interest diverges here and from relativistic QM one usually proceeds to
relativistic Quantum Field Theory. 
Hence it seems that there is a 
problem which most of the books on non-relativistic QM do not mention. 
Considering the importance of the current in interpreting QM, it seems
that it is worthwhile to fill this gap.

To make the problem concrete, 
let us start with the Pauli equation for an electron
in the presence of a electromagnetic field $A_{\mu}$
\begin{equation}
i {\partial \psi \over \partial t}=
\left[{1 \over 2m}(-i \vec{\nabla}-e\vec{A})^2-{e \over 2m}\vec{\sigma}
\cdot {\cal \vec B}+e A_0\right]\psi=H_{Pauli}\psi
\end{equation}  
where $\psi$ is a two component spinor. We have set $\hbar=c=1$.
Although mostly we will be concerned with electromagnetic interactions,
the point we are making (namely the correct form of the current) is
in fact
independent of the detailed form of the interaction. 
We will comment on this later in the text. 
The probability density
\begin{equation}
\rho=\psi^{\dagger}\psi
\end{equation}
only has a consistent, well-defined interpretation 
if it satisfies the
continuity equation
\begin{equation}
{\partial \rho \over \partial t} + \vec{\nabla} \cdot \vec j=0
\end{equation}
Extending the standard  prescription for construction of $\vec j$ from the
Schr\"{o}dinger case (i.e. we use eq.(1) and its hermitian conjugate
in computing $\partial \rho /\partial t $) one finds a current
which we denote here by $\vec{j}'$
\begin{eqnarray}
\vec{j}'&=& -{i \over 2m}\left(\psi^{\dagger}\vec{\nabla}\psi -
(\vec{\nabla}\psi^{\dagger})\psi\right)
-{e \over m}\vec{A}\psi^{\dagger}\psi \nonumber \\
&=& {1 \over m}\Im m(\psi^{\dagger}\vec{\nabla}\psi)-{e \over m}
\vec{A}\rho
\end{eqnarray}
This expression is gauge invariant (thanks to the $\vec A \rho$ term)
and could be,
in principle, a good candidate for the complete quantum mechanical
current of spin-$1/2$ fermions if we could make it plausible that
(4) is in some sense unique. This is, however, not the case. We can
trivially add to $\vec{j}'$ a gauge invariant term proportional to
\begin{equation}
\vec{\nabla}\vec{\times}(\psi^{\dagger}\vec{\sigma}\psi)
\end{equation}
without changing the continuity equation (3). Note that in the Schr\"odinger
case it is not possible to construct a `curl-term' which is
first order in the derivative and second order in the wave functions.

Hence there 
is a priori no way to decide (not only for the electromagnetic interaction), 
from the point of view of non-relativistic QM, 
whether a term like in eq.(5) should be added
to $\vec{j}'$ or not (and if yes
what is the proportionality factor). Since the current is a physical
observable, this apparent ambiguity must have a unique resolution.
Indeed, as will be evident below, there is no
such ambiguity in the full physical theory 
as we can fix the current uniquely by
using relativistic arguments.
Note also
that once this question is settled, the electric current has to be
$\vec{J}=e\vec{j}$.

One could of course argue that only `orbital-terms' like
$\psi^{\dagger}\vec{\nabla}\psi$ should enter 
in $\vec{j}$ and hence also in $\vec{J}$. This has a classical flavour 
and cannot be regarded as a compelling argument. The correct
approach should use a non-relativistic reduction of both, the relativistic
wave equation {\it and its current} and we will see below that the 
above naive
argument does not hold.

Before performing the non-relativistic reduction
for the Dirac equation let us emphasize here two points.
Relativistic QM including external fields 
is well-defined below the particle anti-particle production
threshold which implies that the external fields should not be too strong.
This is to stress the correctness 
of the relativistic external field problem.
While historically one of the first checks of any relativistic theory
has been to test that it yielded the standard non-relativistic limit, by
now the relativistic theory, and in particular here the Dirac equation,
is well-established. Thus when there is a non-relativistic
ambiguity such as we have seen above, we may safely use the relativistic
Dirac theory to find the correct non-relativistic limit.

The Dirac equation in the Dirac represention of the $\gamma_{\mu}$-matrices
reads \cite{bj}
\begin{equation}
i{\partial \over \partial t}\left(\begin{array}{lcr}\psi \\ \chi\end{array}
\right)=
\vec{\sigma}\cdot \vec{\pi}\left(\begin{array}{lcr}\chi \\ \psi \end{array}
\right)
+eA_0 \left(\begin{array}{lcr}\psi \\ \chi \end{array}\right) 
-2m \left(\begin{array}{lcr}0 \\ \chi \end{array}\right)
\end{equation}
where $\vec{\pi}=-i\vec{\nabla}-e\vec{A}$, and $\psi$ and  $\chi$
are both two component spinors. The non-relativistic reduction starts by 
assuming the kinetic energy and field strength to be small compared to the 
mass $m$. Then one of the equations in (6) is approximately
\begin{equation}
\chi \simeq {\vec{\sigma}\cdot \vec{\pi} \over 2m}\psi
\end{equation}
Inserting this in (6) we obtain the Pauli equation (1) for the spinor 
$\psi$.  A similar reduction of the probability density yields
eq.(2) up to terms of order $v^2$ which are of the form
$(1/4m^2)(\vec{\sigma}\cdot\vec{\pi}\psi)^{\dagger} 
(\vec{\sigma}\cdot \vec{\pi}\psi)$.
Of course we should also apply the same approximation to the spatial 
components of the Dirac current
\begin{equation}
\vec{j}_{Dirac}=\Psi^{\dagger}\vec{\alpha}\Psi=-\left(\psi^{\dagger}
\vec{\sigma}\chi +\chi^{\dagger}\vec{\sigma}\psi\right)
\end{equation}
where $\vec{\alpha}=\gamma_0\vec{\gamma}$. Inserting in (8) 
the non-relativistic approximation (7) and using 
$\sigma_i \sigma_j=i\epsilon_{ijk}\sigma_k
+\delta_{ij}$ we find the non-relativistic version of the current
\begin{equation}
\vec{j} = \vec{j}'+{1 \over 2m}\vec{\nabla}\vec{\times}(\psi^{\dagger}
\vec{\sigma}\psi) + {\cal O}(v^2/c^2)
\end{equation}
where $\vec{j}'$ has been already defined in (4). Equation (9) is the 
correct non-relativistic spin-$1/2$ current of the Pauli 
equation. The question about the ambiguity of the Pauli current posed
at the beginning has been completely answered.

We see that the correct 
electric current to be used, say, in addressing questions about 
conductivity indeed contains a 
`spin-term' of the form (5)
as well as the usual `orbital-terms' $\vec{j}'$. 
Some comments are in order here. First note that whereas $\vec{j}'$ 
depends explicitly
on the interaction (potential) used in the Pauli equation (1), the `spin-term'
$(1/2m) \vec{\nabla}\vec{\times}(\psi^{\dagger}\vec{\sigma}\psi)$ does not.
This is clear from eq. (4) where the vector potential enters explicitly.
Had we used an interaction other than the electromagnetic one,
$\vec{j}'$ could then still be constructed as in the Schr\"odinger case,
but the `spin-term' would then follow from a corresponding 
non-relativistic reduction of the Dirac equation (coupled to this interaction)
and its current. In other words this term will always be present, 
regardless of the interaction, 
and indeed even in the interaction-free case (it is even
an easier excercise to perform 
the non-relativistic reduction for free electrons).
From the above it follows that while 
$\vec{j}'$ is closely related to the detailed form
of the Pauli equation, the `spin-term' is not. That is why from the point
of view of non-relativistic quantum mechanics there seems to be an ambiguity. 
It seems that, in general, the non-relativistic result contains less
information when viewed independently from its relativistic `parent'.
This is not surprising as imposing a symmetry, here the Lorentz symmetry, 
limits the number
of choices and can therefore `seal the fate' of a possible structure 
of a term. 
In the case of electromagnetic interactions we can relate the 'spin-term'
to the $(e/2m) \vec{\sigma}\cdot \vec{B}$ term in the Hamiltonian of the Pauli
equation, provided we are allowed to use some arguments from field theory
where a part of the interaction Hamiltonian is given by
$e\int d^3x \vec{j}\cdot\vec{A}$. Inserting here the `spin-term'
of the current (9) we recover, after partial integration,
the $(e/2m) \vec{\sigma}\cdot \vec{B}$ term of the Hamiltonian (we assume
here $\vec{B}$ to be constant and the wave-packets localized). Note that
this argument, in the case of electromagnetic interactions, would
link the Pauli equation with the full current (9) 
and produce the correct result seemingly without 
the non-relativistic reduction. However, there are several drawbacks
to it. For instance, we should not mix (quantum) field theory with 
`point-particle' (quantum) physics. Indeed, in the latter the current
is a functional of the wave function which in turn is the solution
of the Pauli equation. A `confirmation' of the Hamiltonian via
the field theory method can thus be understood as  
a useful heuristic argument.
As mentioned above, the 
presence of the `spin-term' in (9) is also independent of the
form of interaction we use. But the argument given above is limited
to the electromagnetic interaction. It can, however,  always be used 
heuristically to check the correctness of (9).

To show the relevance of the `new' term in eq. (9) 
explicitly, let us solve the Pauli equation (1) 
for an electron in an uniform magnetic field ${\cal \vec{B}}$ (pointing
in the positive $z$-direction) and an uniform electric field ${\cal \vec{E}}$
(in the negative $x$-direction, say). The electromagnetic field configuration
is then like that of the Hall effect \cite{hall}. We are considering here the case
of one electron in the presence of electromagnetic fields, but in vacuum
otherwise. This is then unlike the Hall effect where the electrons
are moving in solids (to avoid confusion, we are not suggesting here a
new way to solve the Quantum Hall effect). 
In the so-called Landau gauge
$A_0={\cal E}x, \,\, A_2={\cal B}x, \,\, A_1=A_3=0, \,\, {\cal E}
=|{\cal \vec{E}}|, \,\, {\cal B}=|{\cal \vec{B}}|$ the Hamiltonian
$H_{Pauli}$ from (1) commutes with $-i\partial /\partial y, \,\, 
-i\partial / \partial z$ and with $\sigma_3$. We make therefore the
ansatz for an unnormalized wave function
\begin{eqnarray}
\psi_{\xi n}&=&e^{-iE_{\xi n}}e^{ip_y y}e^{ip_z z}\Phi_{\xi n}
\nonumber \\
\Phi_{\xi =1, n}&=&\left(\begin{array}{lcr}\varphi_n \\ 0\end{array}
\right), \,\,\,\, \Phi_{\xi =-1,n}=\left(\begin{array}{lcr} 0 \\
\varphi_n \end{array}\right)
\end{eqnarray}
where the quantum number $\xi$ denotes different polarizations.
We leave the wave function unlocalized in the $y$ and $z$ direction.
The eigenvalue problem can be then brought into a form familiar from
the one-dimensional harmonic oscillator
\begin{eqnarray}
\left[E_{\xi n}-{p_y^2 + p_z^2 \over 2m}+ {1 \over 2}
\omega \xi +{1 \over 2}m\omega^2 \beta^2 \right]\varphi_n
\nonumber \\
=(n+ {1 \over 2})\omega \varphi_n=\left[-{ 1 \over 2m}{\partial^2
\over \partial x'^2}+{1 \over 2}m\omega^2 \eta^2\right]
\varphi_n 
\end{eqnarray}
where
\begin{eqnarray}
\omega &=&{e {\cal B} \over m}, \,\,\, \beta=-{e{\cal E}-\omega p_y
\over m \omega^2}\nonumber \\
x'&=&x-\beta, \,\,\, \eta =\sqrt{m\omega}x'
\end{eqnarray}
The energy eigenvalues can be easily shownto be
\begin{equation}
E_{\xi n}=(n+{1 \over 2})\omega -{1 \over 2}\omega \xi +{p_y^2 +p_z^2
\over 2m} -{1 \over 2}m\omega^2 \beta^2
\end{equation}
while the eigenfunctions are given in terms of Hermite polynomials in
the variable $\eta$
\begin{eqnarray}
\varphi_n &=& N_n e^{-{1 \over 2}\eta^2}H_n(\eta)\nonumber \\
N_n &=& \sqrt{\sqrt{{m\omega \over \pi}}{1 \over 2^n n!}}
\end{eqnarray}
We can now compute the components of the electric current 
$\vec{J}=e\vec{j}$ with $\vec{j}$ given in (9). For definiteness
we do it for $\xi =1$.
\begin{eqnarray}
J_x &=& {e \over m}{\partial \over \partial y}\rho_n=0
\nonumber \\
J_z &=& e {p_z \over m} \rho_n \nonumber \\
J_y &=& e\rho_n {{\cal E} \over {\cal B}} -
{e \over m} 2n e^{-\eta^2}H_n(\eta)H_{n-1}(\eta)
\end{eqnarray}
where $\rho_n= \psi^{\dagger}_{\xi=1,n}\psi_{\xi=1, n}$. 
For $n=0$, i.e. in the ground state, the second term in $J_y$ vanishes. 
The first term, viz.
$e\rho_n {\cal E}/{\cal B}$, corresponds in this form to the classical result
(see e.g. \cite{hall}). In the ground state therefore, the classical 
and quantum mechanical results coincide.  For higher excited states there is, 
however, a new contribution to $J_y$ (proportional to $2ne/m$) which is of 
purely
quantum mechanical origin and 
which can be traced back to the `spin-term' in eq.(5).
As already stated, we have not tried to solve here the Quantum Hall effect \cite{qhall}.
Our main motivation was to point out the relevance of the `spin-term'
to the (electric) current. It would clearly be interesting to study the
effects of the extra term in the current in various physical applications. 
We feel it is reasonable to speculate that
this term might play a role in
problems concerning conductivity in solids.
Finally, we recall that the proof of the need for
such a term in the non-relativistic
current was decided essentially by relativistic QM. This is one of the few 
places where relativistic QM can resolve a problem of the non-relativistic 
theory. We think that text books on non-relativistic QM should include
at least a note on the different  
(as compared to the Schr\"odinger case) nature of the quantum mechanical
Pauli current in order not to give the impression that in constructing 
the Pauli current it suffices to copy the steps from the Schr\"odinger
case. The quantum mechanical current is not only important for
the correct interpretation of QM, but also in calculating conductivity
etc. in solid matter. In teaching this subject, one can 
point out the power of symmetry arguments which limits our choice of possible
terms. Since the appearance of the ambiguity 
(from the non-relativistic point of view) in the Pauli current is
physically not acceptable (the current is an observable), this 
hints towards the need of a more general theory which includes
relativity. In spite of the fact that in practical calculations
concerning properties of solids relativity does not play a big part,  
it plays a role, as shown above, on a more fundamental level.
This is then also a lesson on the unity of physics. 

\vskip 1cm
{\bf Acknowledgments}. 
I would like to thank R. Godbole and M. Lavelle for many
valuable discussions and suggestions.
This work has been 
supported by the Spanish Ministerio de Educacion y Ciencia.


\vskip 3cm
\begin{thebibliography}{99}
\bibitem{one}
A disclaimer is in order at this point.
The author of the present paper does not claim that there do not exist
any books with a discussion of the spin-$1/2$ current. It is then possible that a discussion can be found where the results are similar or opposite 
to these presented here . 
If so, the exercise of this note here is to point out
the problem and its solution.
\bibitem{two}
The author noticed the problem of the Pauli current while working
on a different but related problem of constructing a current for
neutral spin-$0$ $K^0$ and $\bar{K^0}$ mesons. There 
one of the problems is to have a relativistic well-defined current.See, 
B. Ancochea, A. Bramon, R. Munoz-Tapia, M. Nowakowski,
``Space-dependent probabilities for $K^0-\bar{K^0}$ oscillations'',
Phys. Lett. {\bf B389}, 149-156 (1996)
\bibitem{bj}
J. D. Bjorken and S. D. Drell, {\it Relativistic Quantum Mechanics},
(McGraw-Hill, 1964), pp. 1-10 
\bibitem{hall} 
For an overview see
in A. P. Balachandran, E. Ercolesi, G. Morandi and A. M. Srivastava,
{\it Hubbard Model and Anyon Superconductivity}, (World Scientific, 1990),
pp. 121-135
\bibitem{qhall}
For a general reference 
see {\it The Quantum Hall Effect}, eds. R. E. Prange and
S. M. Girvin, (Springer-Verlag, 1990) 2nd edition; 
also G. Morandi, {\it Quantum Hall Effect},
Bibliopolis 1988 
\end{thebibliography}
\end{document}